\documentclass[12pt]{article}
\usepackage{mathtext}
\usepackage{cite}
\usepackage[english]{babel}
\usepackage{amsmath}
\usepackage{amsfonts}
\usepackage{textcomp}
\usepackage{graphicx}
\title{Thermodynamics and kinetics of boundary friction}
\author{I.A.~Lyashenko, A.V.~Khomenko, L.S.~Metlov$^\ast$ \\
Sumy State University, 40007 Sumy, Ukraine\\
$^\ast$Donetsk Institute for Physics and Engineering named after A.A. Galkin \\
of the NASU, 83114, Donetsk, Ukraine\\
E-mails: nabla04@ukr.net, khom@mss.sumdu.edu.ua, lsmet@fti.dn.ua 
}
\date{\ }
\begin{document}
\maketitle
\thispagestyle{empty}
\def\abstractname{\ }
\begin{abstract}
\vspace{-1.8cm}
A deterministic theory describing the behavior of an ultrathin lubricant film between two atomically-smooth solid surfaces is proposed.
For the description of lubricant state the parameter of excess volume arising due to chaotization of solid medium structure in the
course of melting is introduced. Thermodynamic and shear melting is described consistently. Dependences of friction force
on temperature of lubricant, shear velocity of rubbing surfaces, and pressure upon surfaces are analyzed.
Within the framework of a simple tribological model the stick-slip mode of friction, when the lubricant periodically melts and
solidifies, is described. The obtained results are qualitatively compared with the experimental data.

{\bf Keywords:} ultrathin lubricant film, friction force, phase transition, stick-slip motion.

PACS  05.70.Ce; 05.70.Ln; 47.15.gm; 62.20.Qp; 64.60.-i; 68.35.Af; 68.60.-p
\end{abstract}

\section{Introduction}\label{sec1}

The boundary friction mode arising with lubricant thickness less than 10 atomic layers is widely investigated
 so far \cite {Persson}. Experiments show that a thin layer of lubricant shows anomalous properties in comparison with volume
lubricants \cite{Yosh}. In particular, the intermittent motion ($stick-slip$) inherent in a dry friction is observed \cite{SRC,Yosh}.
Such mode is explained as the solidification caused by compression of rubbing surfaces followed by jump-like melting at a shear stress
larger than the yield limit (''shear melting``).

There are some phenomenological models, for example, thermodynamic \cite{Popov}, mechanistic \cite{Carlson} and synergetic \cite{KhYu}
ones, allowing to explain the experimental results partially. They are either deterministic \cite{KhYu,Carlson} or
stochastic \cite{Filippov,Filippov1} in nature. Studies by using  molecular dynamics methods are also known
\cite{Braun,prd,Khome2010, SS_2010}. It turns out that the lubricant provides several kinetic modes with transitions leading to the
$stick-slip$ friction \cite {Yosh}. Such transitions are the first-order phase transitions \cite{land_inst} however between the states
which are not true thermodynamic phases but the  kinetic modes of friction.
In work \cite{Filippov} three modes are found: sliding at low shear velocities, regular $stick-slip$ mode, and sliding at the high
shear velocities. Their existence has been verified in numerous experiments \cite{Persson,Yosh,SRC,silica}.

In work \cite{KhYu}, within the framework of Lorenz model, to approximate viscoelastic medium, an approach has been developed according
to which the transition of ultrathin lubricant film from solid- into liquid-like state results from the thermodynamic and shear melting.
The processes due to self-organization of shear stress and strain fields and of lubricant temperature are described in view of
additive noises of the specified quantities \cite{JTPh,dissipative,fnl_10} and correlated temperature fluctuations \cite{uhl_fnl}.
The reasons for jump-like melting and hysteresis, observed experimentally \cite{exp1,exp_n,Isr_rev}, are considered in
works \cite{FTT,PhysLettA}. Here, realization conditions of these features are determined with the deformation defect of the shear
modulus taken into account. In work \cite{period}, within the framework of the mentioned model the periodic $stick-slip$ mode of
friction is described which,  however, has a stochastic component and can be realized only in the presence of fluctuations. The
drawback of the model is that it does not consider the load applied to the surfaces of friction and that basic equations are derived
by several approximations in \cite{KhYu}.

In the present work, the theory based on the expansion of free energy of system into a power series of parameter $f$ which, being
squared, represents the excess volume arising from the formation of defect structure in the lubricant during melting. The liquid-like state is usually interpreted as a section of plastic flow in the loading diagram and characterized by the
presence of defects in the lubricant \cite{Popov}. Here, the approach based on the Landau theory of phase transitions is
used \cite{PREMet,Ran,MFNT,MFNT_Metlov,preprint} to describe strongly non-equilibrium processes, occurring at sliding of two rubbing
solid surfaces separated by a layer of lubricant.

\section{Basic equations}\label{sec2}

The melting of lubricants whose thickness is less than 10 molecular layers  is judged by the increase in their volume \cite{Braun}
and diffusion factor \cite{Braun,prd,liqtosol,thompson}. Since, of the two, the experimentally observable quantity is the volume, to
describe the lubricant state we introduce parameter $f$, which, when squared, stands for the excess volume arising due to chaotization
of structure of a solid medium under of melting. With the increase in $f$ magnitude the density of defects in the lubricant is growing,
and owing to their transport under the applied stresses the lubricant passes to a kinetic mode of plastic flow (liquid-like phase).

Let's write down the dependence of the density of free energy $\Phi$ on excess volume in the form of expansion in terms of parameter $f$
\begin{eqnarray}
\Phi &=& \Phi_0 + \frac{c}{2}\left({\nabla f}\right)^2 + \varphi_0 f^2 - \frac{1}{4}\varphi_1 f^4 + \frac{1}{6}\varphi_2 f^6,
\label{int_energy}
\end{eqnarray}
where $\Phi_0$, $c$, $\varphi_0$, $\varphi_1$, $\varphi_2$ are expansion constants.

We consider the dependences of elastic strains $\varepsilon_{ij}^e$ and lubricant temperature $T$ on invariants for the smaller
expansion powers only
\begin{eqnarray}
\Phi_0 &=& \Phi_0^\ast + \frac{1}{2}\lambda\left(\varepsilon_{ii}^e\right)^2 + \mu\left(\varepsilon_{ij}^e\right)^2, \\
\varphi_0 &=& \varphi_0^\ast - \frac{1}{2}\bar\lambda\left(\varepsilon_{ii}^e\right)^2 -
\bar\mu\left(\varepsilon _{ij}^e \right)^2 - \alpha T.
\label{varphi_const}
\end{eqnarray}
The first invariant represents the trace of strain tensor $\varepsilon_{ii}^e = \varepsilon_1^e+\varepsilon_2^e+\varepsilon_3^e$
and the second one is defined by the expression \cite{Kachanov}
\begin{equation}
(\varepsilon_{ij}^e)^2 \equiv (\varepsilon_{ll}^e)^2 - 2I_2 =
(\varepsilon_1^e+\varepsilon_2^e+\varepsilon_3^e)^2- 2(\varepsilon_1^e\varepsilon_2^e+\varepsilon_1^e\varepsilon_3^e+
\varepsilon_2^e\varepsilon_3^e) =
(\varepsilon_1^e)^2+(\varepsilon_2^e)^2+(\varepsilon_3^e)^2.
\end{equation}

According to (\ref{int_energy}), in the lubricant the elastic stresses appear 
\begin{equation}
\sigma_{ij}^e = \frac{\partial\Phi}{\partial\varepsilon_{ij}^e} =
\lambda\varepsilon_{ii}^e\delta_{ij} + 2\mu \varepsilon_{ij}^e - \left(\bar\lambda\varepsilon_{ii}^e \delta_{ij}
+2\bar\mu\varepsilon_{ij}^e\right)f^2.
\label{sigma_ij}
\end{equation}
Equation (\ref{sigma_ij}) can take the form of the effective Hooke law
\begin{equation}
\sigma_{ij}^e = 2\mu_{eff}\varepsilon_{ij}^e + \lambda_{eff}\varepsilon_{ii}^e\delta_{ij}
\label{hooke}
\end{equation}
with the effective elastic parameters\footnote{For $f^2>\mu/\bar\mu$, we put $\mu_{eff}=0$ and for $f^2>\lambda/\bar\lambda$,
we take $\lambda_{eff}=0$.}
\begin{eqnarray}
\mu_{eff} &=& \mu - \bar\mu f^2,\label{mu_eff}\\
\lambda_{eff} &=& \lambda - \bar\lambda f^2,
\label{lambda_eff}
\end{eqnarray}
which decrease with melting for increasing $f$.

It is easy to show that
\begin{eqnarray}
\varepsilon_{ii}^e &=& \frac{n}{\lambda_{eff}+\mu_{eff}}, \label{first_inv}\\
(\varepsilon_{ij}^e)^2 &=&
\frac{1}{2}\left[\left(\frac{\tau}{\mu_{eff}}\right)^2+\left(\varepsilon_{ii}^e\right)^2\right],
\label{second_inv}
\end{eqnarray}
where $n$, $\tau$ are normal and shear components of the stresses acting on lubricant from the side of rubbing
surfaces\footnote{Shear stress $\tau$ is defined from expression (\ref{hooke}) for $i\ne j$, i.e. $\delta_{ij}=0$.
When $\mu_{eff}=0$ the term $\tau/\mu_{eff}$ in (\ref{second_inv}), should be substituted by $2\varepsilon_{ij}^e$,
according to (\ref{hooke}).}. Relationships (\ref{first_inv}), (\ref{second_inv}) represent the relationship between components of
tensors and their invariants of the linear theory of elasticity \cite{Kachanov}.

Let's write down the kinetic equation for non-equilibrium parameter $f$ in the form of the Landau-Khalatnikov equation
\begin{equation}
\tau_f\dot f = -\frac{\partial\Phi}{\partial f},
\label{newton}
\end{equation}
where relaxation time $\tau_f$ is introduced. In the explicit form it is written as
\begin{equation}
\tau_f\frac{\partial f}{\partial t} = -c\nabla^2f - 2\varphi_0 f + \varphi_1 f^3 - \varphi_2 f^5 -
\frac{2n^2\left(\bar\lambda+\bar\mu\right)f}{\left(\lambda_{eff}+\mu_{eff}\right)^2},
\label{kin_h}
\end{equation}
with the last term showing the dependence of (\ref{first_inv}) and (\ref{second_inv})  on $f$.

We derive the equation connecting the relative velocity of shear for rubbing surfaces $V_{ij}$ and the elastic strains arising in
the lubricant $\varepsilon_{ij}^e$. For this purpose we use the Debye approximation relating the elastic strain and the plastic
one $\varepsilon_{ij}^{pl}$ \cite{Popov}:
\begin{equation}
\dot\varepsilon_{ij}^{pl} = \frac{\varepsilon_{ij}^e}{\tau_\varepsilon},
\label{e_pl}
\end{equation}
where $\tau_\varepsilon$ is the Maxwell relaxation time of internal stress. The total strain in a layer is defined as
\begin{equation}
\varepsilon_{ij} = \varepsilon_{ij}^{e} + \varepsilon_{ij}^{pl}.
\end{equation}
This strain fixes the velocity of motion of the top block according to equation \cite{Wear}
\begin{equation}
V_{ij}=h\dot\varepsilon_{ij} = h (\dot\varepsilon_{ij}^e + \dot\varepsilon_{ij}^{pl}),
\label{V}
\end{equation}
where $h$ is the thickness of lubricant film. The last three relationships give the expression for elastic components of shear strain:
\begin{equation}
\tau_\varepsilon\dot\varepsilon_{ij}^e = -\varepsilon_{ij}^e + \frac{V_{ij}\tau_\varepsilon}{h}.
\label{e_ij_solid}
\end{equation}
Further, for simplicity, a homogeneous system is considered, and in Eqs.~(\ref{int_energy}), (\ref{kin_h}) we put $\nabla \equiv 0$.

\section{Friction force}

In view of definitions (\ref{varphi_const}), (\ref{hooke}) - (\ref{second_inv}) the set of kinetic equations (\ref{kin_h}),
(\ref{e_ij_solid}) is closed and can be used for studying the kinetics of lubricant melting. In this section we consider the
stationary modes of friction resulting from system evolution. According to (\ref{e_ij_solid}), in the course of time, the stationary
value of elastic component of shear strain is set in
\begin{equation}
\varepsilon_{ij0}^{e} = \frac{V_{ij}\tau_\varepsilon}{h}.
\label{st}
\end{equation}
For finding the stationary states of all quantities it is necessary to solve numerically the evolution equation (\ref{kin_h}),
using (\ref{varphi_const}), (\ref{hooke}) -- (\ref{second_inv}) and determining the value of strain from (\ref{st}).

In experimental studies, the dependences of friction force on velocity of shear, thickness of lubricant layers, and load on the
rubbing surfaces are often determined \cite{Persson,Yosh,SRC,exp1,exp_n,Isr_rev}. Let's analyze how the lubricant temperature and
velocity of shear affect the friction force.

In the course of friction, both elastic $\sigma_{ij}^e$ and viscous $\sigma_{ij}^{visc}$ stresses arise in the lubricant. The total
stress in a layer is the sum of these two components
\begin{equation}
\sigma_{ij} = \sigma_{ij}^e + \sigma_{ij}^{visc}.
\label{sigma_all}
\end{equation}
The total friction force is defined in a standard way:
\begin{equation}
F_{ij} = \sigma_{ij}A,
\label{F_begin}
\end{equation}
where $A$ is the area of contacting surfaces. Viscous stresses in a layer are given by the formula \cite{Wear}
\begin{equation}
\sigma_{ij}^{visc} = \frac{\eta_{eff}V_{ij}}{h},
\label{sigma_ij_v_begin}
\end{equation}
where $\eta_{eff}$ is the effective viscosity measured only experimentally, and in the boundary mode
\begin{equation}
\eta_{eff} \sim \left(\dot\varepsilon_{ij}\right)^\gamma,
\label{eta_eff}
\end{equation}
and for the majority of systems $ \gamma=2/3$. In view of (\ref{V}), (\ref{eta_eff}) the expression
for viscous stresses (\ref{sigma_ij_v_begin}) is written in the form:
\begin{equation}
\sigma_{ij}^{visc} = \left(\frac{V_{ij}}{h}\right)^{\gamma+1}.
\label{sigma_ij_v}
\end{equation}
After substitution of (\ref{sigma_all}) and (\ref{sigma_ij_v}) in (\ref{F_begin}) we have the required expression for friction
force\footnote{Here the sign function $\rm sgn(x)$ and absolute value of shear velocity $|V_{ij}|$ are introduced, since $V_{ij}$
can accept negative values too.}:
\begin{equation}
F_{ij} = \left[\sigma_{ij}^e + {\rm sgn}(V_{ij})\left(\frac{|V_{ij}|}{h}\right)^{\gamma+1}\right] A,
\label{F}
\end{equation}
where $\sigma_{ij}^e$ is defined by the formula (\ref{hooke}) for $i\ne j$.

In experiments, the friction surfaces are often the atomically-smooth surfaces of mica, and lubricants are the quasispherical molecules
of octamethylcyclotetrasiloxane (OMCTS) and linear chain molecules of tetradecane or hexadecane \cite{Yosh,Isr_rev}. The experimental
conditions are as follows: the thickness of lubricant $h\sim 10 ^ {-9} $~m, the contact area $A\sim 3\cdot 10^{-9}$~m$^2$, load on
the top surface of friction $L = (2\div 60) \cdot 10^{-3}$~N that corresponds to normal
stresses $n =-L/A = - (6.67\div 200) \cdot 10^{5}$~Pa. The friction force $F\sim (2\div 40) \cdot 10^{-3}$~N. In the mentioned
experiments it has been found that the lubricant melts at a temperature above the critical value $T>T_{c0} \sim 300$~K, or at a
velocity of shear $V>V_c\sim 400$~nm/s. These values may much differ depending on lubricant and geometry of the experiment.

According to experimental data, the following values of the constants of the theory are chosen for the considered model:
$\Phi_0^\ast = 20~{\rm J/m^3},
\lambda = 2\cdot 10^{11}~{\rm Pa},
\bar\lambda = 10^{8}~{\rm Pa},
\mu = 4.1\cdot 10^{11}~{\rm Pa},
\bar\mu = 4\cdot 10^{11}~{\rm Pa},
\varphi_0^\ast = 185~{\rm J/m^3},
\varphi_1 = 570~{\rm J/m^3},
\varphi_2 = 3200~{\rm J/m^3},
\alpha = 0.6~{\rm J\cdot K^{-1}/m^{3}},
h = 10^{-9}~{\rm m},
\tau_f = 1~{\rm Pa{\cdot}s},
\tau_\varepsilon = 10^{-8}~{\rm s}.$
Note that the relaxation time of excess volume $\tau_f$ has the dimension of viscosity. In fact, this means that the time of reaching
the stationary friction mode increases with the growth of the effective viscosity of the lubricant.

\begin{figure}[!ht]
\hspace{-1cm}
\centerline{\includegraphics[width=140mm]{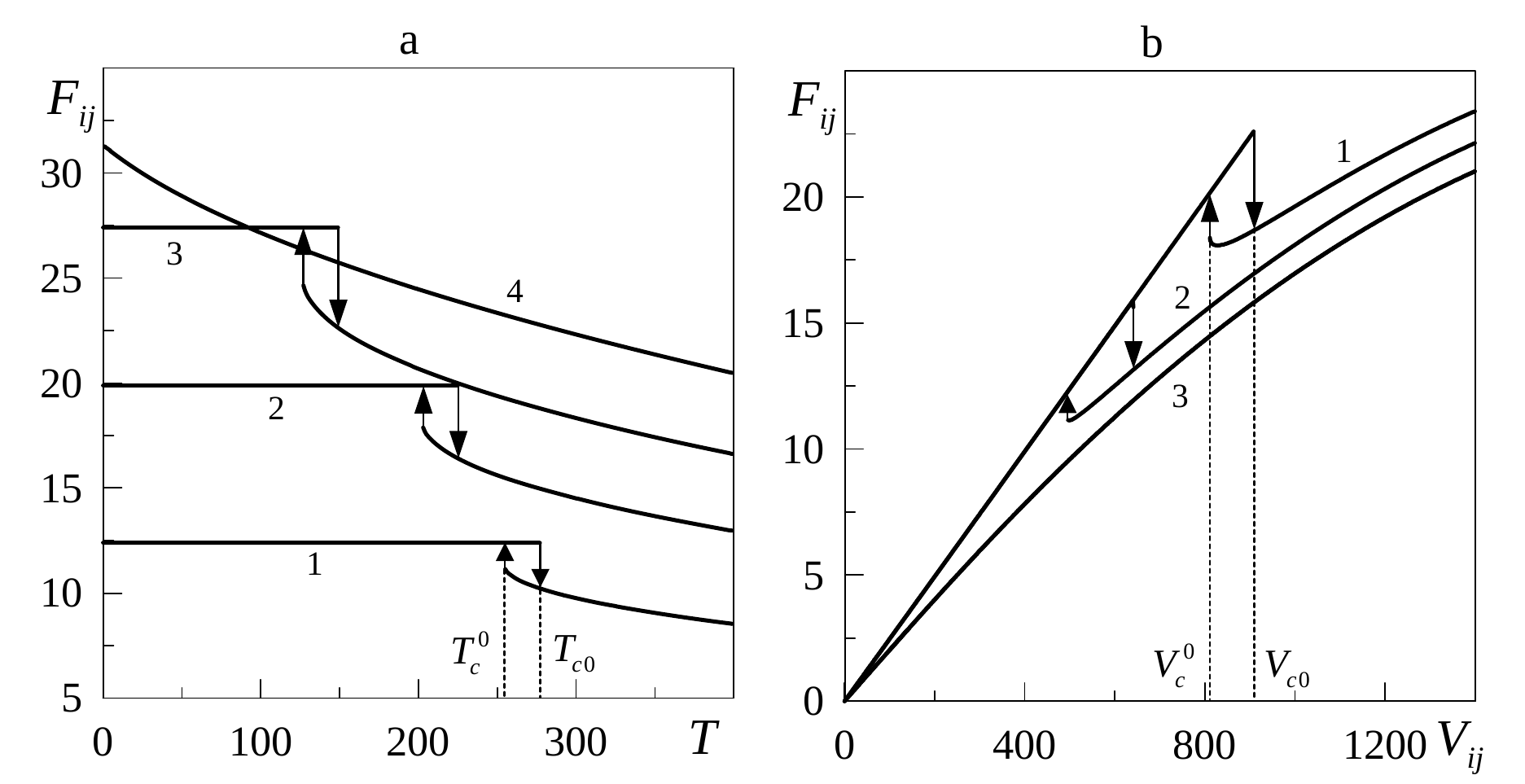}}
\caption{Dependence of the stationary value of the total friction force $F_{ij}$~(mN) (\ref{F}) on the temperature of lubricant $T$~(K) and the 
velocity of shear $V_{ij}$~(nm/s) for $\gamma=2/3$, $A=3\cdot 10^{-9}$~m$^2$, $n =-7\cdot 10^5~{\rm Pa}$:
a) curves 1--4 correspond to the constant values of velocities of shear $V_{ij}$=500~nm/s, 800~nm/s, 1100~nm/s, 1550~nm/s;
b) curves 1--3 correspond to the fixed values of temperature $T$=200~K, 255~K, 310~K.}
\label{fig1}
\end{figure}

Figure~\ref{fig1}a illustrates the fact that the friction force decreases with lubricant temperature increase.
Let's analyze the curve~1 in detail. At first, with temperature increase to point of phase transition at $T<T_{c0} $ the excess volume
is equal to zero, therefore the effective shear modulus $2\mu_{eff}$ (\ref{mu_eff}) takes constant value. Since the shear velocity is
invariable, in accordance with (\ref{st}), the constant value of strain is realized, leading to constant shear stress (\ref{hooke}).
Thus, the both components in (\ref{F}) do not change with temperature increase to a point of phase transition, and the friction force
is constant. When the temperature becomes higher than the critical value $T>T_{c0}$, the value of excess volume $f^2$ increases abruptly,
and the lubricant melts that leads to a sharp decrease in the total friction force.

If after melting the temperature is increased, the excess volume increases resulting in reduction of the shear modulus and, as a
consequence, in a smaller value of the total friction force. With temperature fall the lubricant solidifies when $T=T_c^0$. Thus, the
dependence is of hysteretic character that corresponds to the first-order phase transitions. With the increase in velocity of shear the
lubricant melts at a smaller value of temperature. When the velocity exceeds the critical value, the lubricant is always liquid-like
independent of temperature (curve 4), and the friction force decreases with temperature increase owing to the decrease in the shear
modulus (the lubricant becomes more liquid).

Thus, at small temperatures ($T<T_c^0$) there is one zero minimum of the potential $\Phi(f)$ (solid-like lubricant). In the range of
temperatures $T_c^0<T<T_{c0} $, together with the specified minimum, two symmetric nonzero minima of $\Phi(f)$
coexist\footnote{Since $f^2$ is the observable quantity, these minima are equivalent in relation to friction mode.},
however, the system can not pass to the state corresponding to those minima since they are separated by maxima with the zero minimum.
At further increase in temperature $T>T_{c0}$ the separating maxima disappear, and according to the mechanism of the first-order phase
transition the lubricant passes to the state corresponding to the minimum of potential with $f\ne 0$, i.e. it melts. If now the
temperature is decreased, there appears a zero minimum and again the system can't pass to the respective state because of the presence
of separating maxima. With their disappearance at $T=T_c^0$ the lubricant solidifies abruptly.

Figure~\ref{fig1}b shows a somewhat different behavior. Here, in accordance with (\ref{F}), at small shear velocities the lubricant is
solid-like, and since $f=0$ the value of shear modulus is maximal. In such a mode an increase in the velocity increases the friction
force (\ref{F}) due to the increasing viscous stresses. At $V>V_{c0}$ the lubricant melts, and elastic shear stresses (\ref{hooke})
decrease sharply and the total friction force decreases in a jump-like manner. With further increase in the velocity, the $F_{ij}$
increases owing to viscous component of the friction force, which is always increasing with the velocity. According to curve~3, in the
liquid-like state the friction force (\ref{F}) increases due to the increase in the velocity. With temperature increase  the lubricant
melts at lower shear velocities. Note, that the results of Fig.~\ref{fig1}b qualitatively coincide with the new map of friction for a
boundary mode proposed in work \cite{Wear} as a result of generalization of experimental data. At present, the dependences of friction
force on temperature similar to those presented in fig.~1a are not measured experimentally.

\section{Stick-slip mode}

\begin{figure}[!ht]
\hspace{-1cm} \centerline{\includegraphics[width=120mm]{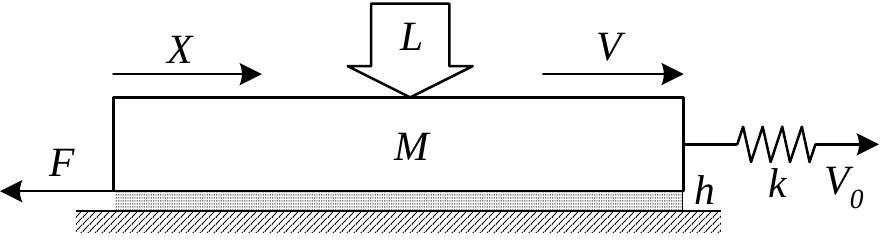}}
\caption{The scheme of tribological system.}
\label{fig2}
\end{figure}

The dependences of figure~\ref{fig1} have been obtained for the fixed shear velocity of the top rubbing surface. However, the dynamic
characteristics of tribological systems are defined by both the friction force, shown in the specified figure, and the properties of
system as a whole. In particular, according to experiments, in the hysteresis region of the dependence in fig.~\ref{fig1}, the
interrupted mode of friction ($stick-slip$) \cite{Yosh,Carlson,Filippov,Filippov1,Braun,Isr_rev,liqtosol,Wear} can be realized. The
present work is devoted to definition of its features. The typical scheme of tribological system is presented in fig.~\ref{fig2}.
Here, the spring of rigidity $k$ is connected to the block of mass $M$ to which additional normal load $L$ is applied. The block is
located on a smooth surface from which it is separated by a layer of lubricant of thickness $h$. The free end of the spring is brought
in motion with a constant velocity $V_0$. Block motion initiates friction force $F$ (\ref{F}) that resists its displacement. Generally,
for ultrathin layers of lubricants in a mode of boundary friction the velocities of block $V$ and spring $V_0$ do not coincide because
of the oscillating character of force $F$ leading to interrupted motion of the block. Such mode resembles dry friction without a
lubricant.

The equation of motion of the top block looks like \cite{Yosh, Popov, Carlson}\footnote{Further, for convenience we omit lower tensor
designations since shear in one direction is considered.}
\begin{equation}
M\ddot X = k\left(V_0 t-X\right)-F.
\label{Move}
\end{equation}
To calculate the time evolution of the friction force, the last equation is to be solved together with (\ref{kin_h}),
(\ref{e_ij_solid}), and the friction force defined from (\ref{F}). However, the relaxation time of strain $\tau_\varepsilon$, may be put
small in comparison with that of excess volume $\tau_f$  owing to lubricant's thinness. Therefore, within the limits of
approximation $\tau_f\gg\tau_\varepsilon$, the two equations (\ref{Move}), (\ref{kin_h}) are solved jointly and the strain is defined
from (\ref{st}).

\begin{figure}[!ht]
\hspace{-1cm} \centerline{\includegraphics[width=120mm]{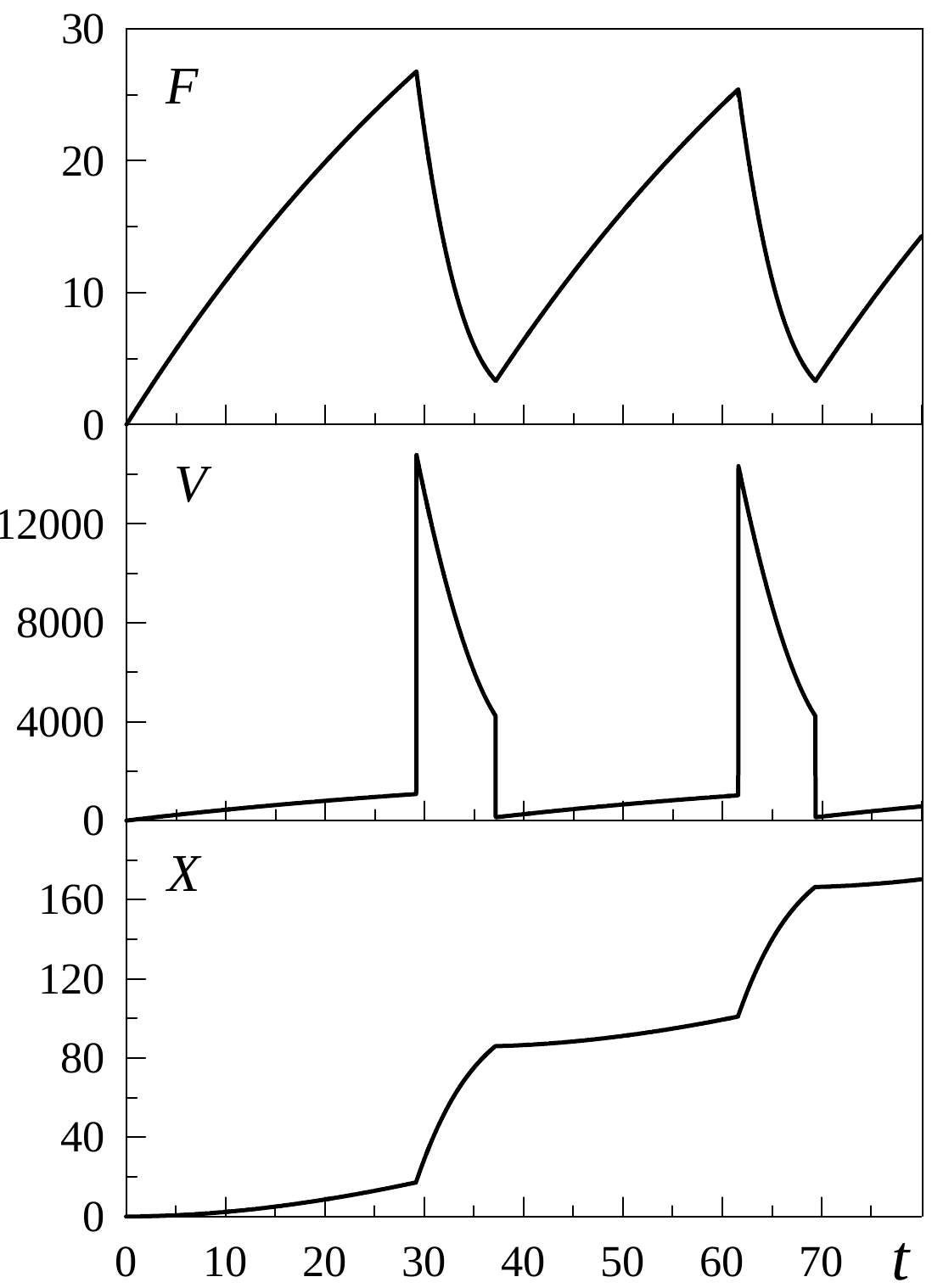}}
\caption{Dependence of the friction force $F$~(mN), the shear velocity of rubbing surface $V$~(nm/s),
and its coordinate $X$~($\mu$m) on the time $t$~(s) for $n=-7\cdot 10^5$~Pa, $M=0.4$~kg, $k=480$~N/m, $T=250$~K, $V_0=2500$~nm/s.}
\label{fig3}
\end{figure}

The result of solution of the specified equations is shown in fig.~\ref{fig3}.
According to the figure, at first, the friction force monotonously increases, since the lubricant is solid-like, and shear velocity $V$
increases. When the velocity exceeds critical value $V_{c0}$, the lubricant melts and the friction force decreases, the shear velocity
of the rubbing block increases, and it quickly moves to a large distance. As a result, the tension of spring and the velocity of shear
decrease. When the velocity becomes less than that for the lubricant to be maintained in liquid-like state, it solidifies, and the
friction force starts again increasing. The described process is periodically recurrent. It is worth noting that the velocity, with
which the lubricant solidifies, does not coincide with the similar velocity shown in fig.~\ref{fig1}. This is because of the sharp
increase in the velocity of shear $V$ at melting and corresponding increase in excess volume $f^2$. According to (\ref{mu_eff}), in
this case the shear modulus becomes less than zero, but it need be considered zero and this deforms the potential
form (\ref{int_energy}). Thus, in the presence of elastic strain (\ref{st}), according to (\ref{hooke}), the elastic stresses in the
lubricant are equal to zero. As a result, the friction force decreases, and the lubricant flows.

\begin{figure}[!ht]
\hspace{-1cm} \centerline{\includegraphics[width=120mm]{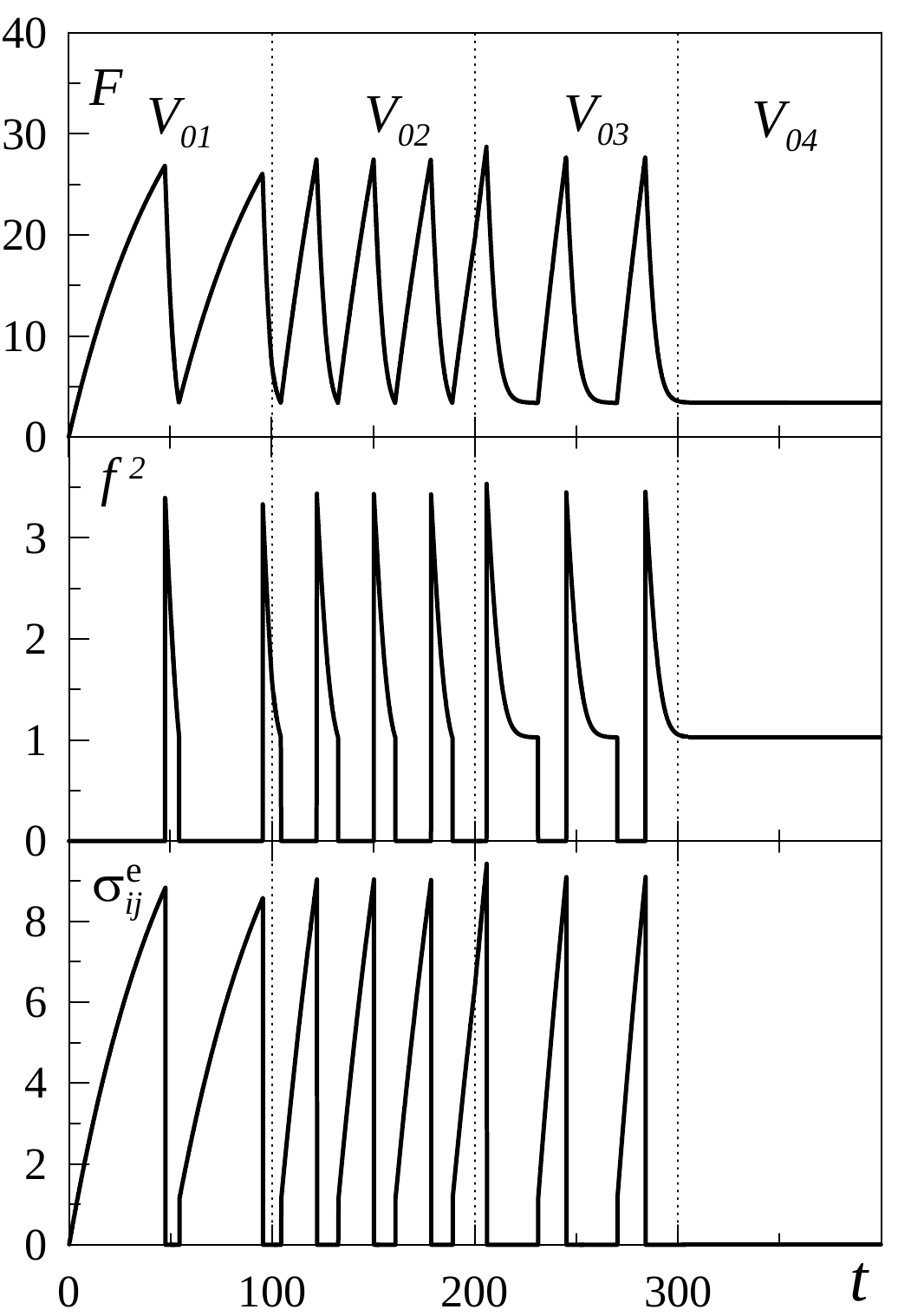}}
\caption{Dependence of the friction force $F$~(mN), the excess volume $f^2$, and the elastic components of shear stress $\sigma_{ij}^e$~(MPa)
on the time $t$~(s) for parameters of fig.~\ref{fig3}, normal stress $n=-30\cdot 10^5$~Pa, and velocities of shear $V_{01}=1800$~nm/s,
$V_{02}=3500$~nm/s, $V_{03}=4286$~nm/s, $V_{04}=4290$~nm/s.}
\label{fig4}
\end{figure}

In fig.~\ref {fig4} there are dependences of the total friction force $F$ (\ref{F}), excess volume $f^2$, and elastic component of
the shear stress $\sigma_{ij}^e$ (\ref{hooke}) on time under velocity $V_0$ increase.
Initially, the movement of the top sheared block ($V_0=V_{01}$) leads to the increase in friction force at zero $f^2$. When the elastic
shear stresses reach the critical value, the shear melting of lubricant is initiated by the first-order phase transition mechanism.
Now, parameter $f^2$ increases abruptly, and elastic stresses become equal to zero. And again, there follows lubricant solidification
since the relative shear velocity of the friction surfaces decreases (see~fig.~3). When it solidifies completely, the elastic stresses
appear there. And again, subsequent increase in the stresses leads to the increase in parameter $f^2$ until it reaches the critical
value needed for melting, and the process is repeated. As a result, the periodic interrupted ($stick-slip$) mode of
melting/solidification becomes steady.
With velocity increase to $V_0=V_{02}$ the frequency of $stick-slip$ peaks increases because at this velocity the critical value of
stresses is reached more rapidly. Accordingly, the lubricant melts more rapidly and in the same interval of time the system succeeds
in making a greater quantity of melting/solidification transitions.
With a higher increase in velocity $V_0=V_{03}$, the frequency of $stick-slip$ peaks decreases again due to the appearance of long
kinetic sections $F=\rm const$ in the $F(t)$ dependence. Note that in this mode, at melting, the excess volume $f^2$  initially sharply
increases owing to rapid increase in shear velocity of the top rubbing block $V$. A smaller value of excess volume $f^2$ reached after
abrupt initial slip of the rubbing surface at the expense of released mechanical potential energy of compressed spring corresponds to
the stationary kinetic section.
With further increase in velocity $V_0=V_{04}$, the interrupted mode disappears, and the kinetic mode of liquid-like lubricant friction
is reached, it is characterized by nonzero value of excess volume $f^2$ and elastic stresses $\sigma_{ij}^e=0$.
Thus, with the increase in velocity, the frequency of $stick-slip$ peaks first increases, then it decreases due to the appearance of
long kinetic sections, and when value of velocity exceeds $V_0$ the $stick-slip$ mode disappears.
The described behavior is in a good agreement with experimental data \cite{Yosh}.

The influence of the external pressure, applied perpendicularly to the friction surfaces, on character of lubricant melting is often
studied experimentally \cite{Yosh,liqtosol}. Such experiments show that pressure effect on the parameters of tribological systems is
not trivial. For example, for a lubricant consisting of chain molecules of hexadecane the critical velocity of shear decreases with
pressure growth, and for spherical molecules of OMCTS it increases, on the contrary \cite{Yosh}. Pressure influences the frequency
and amplitude of $stick-slip$ transitions too \cite{Yosh}. Within the limits of our model, according to equation (\ref{kin_h}), the
increase in load on the friction surface reduces the excess volume that should promote lubricant solidification.

\begin{figure}[!ht]
\hspace{-1cm} \centerline{\includegraphics[width=120mm]{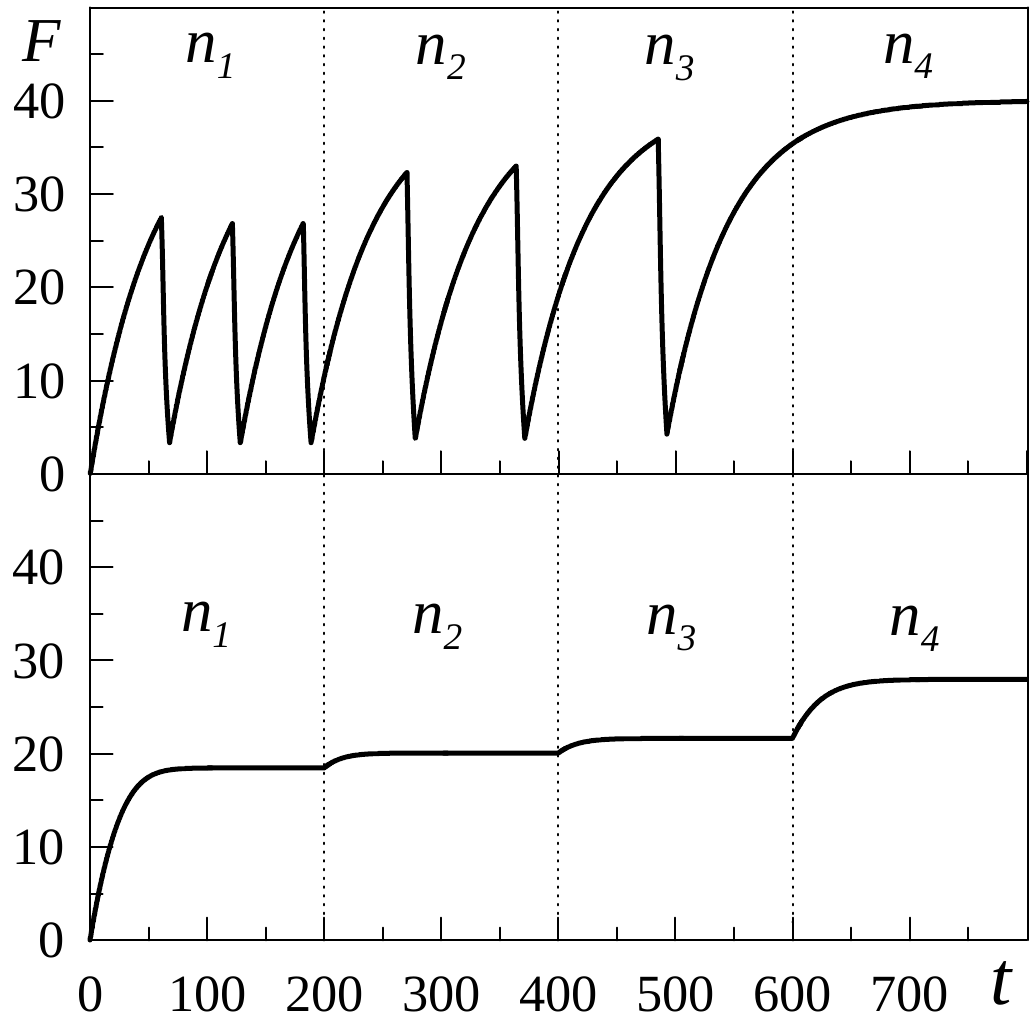}}
\caption{Dependence of the friction force $F$~(mN) on the time $t$~(s) for parameters of fig.~\ref{fig3}, shear velocity $V_0=1600$~nm/s,
and external normal load $n_1=-7\cdot 10^5$~Pa, $n_2=-70\cdot 10^5$~Pa, $n_3=-100\cdot 10^5$~Pa, $n_4=-180\cdot 10^5$~Pa.
The top panel corresponds to the temperature $T=235$~K, the bottom one to $T=540$~K.}
\label{fig5}
\end{figure}

In figure~\ref {fig5}, the time dependence of friction force is shown at various values of pressure acting to compress friction surfaces.
At a temperature below the critical value (the top panel of figure) the interrupted mode of friction is realized, and with pressure
growth the amplitude of $stick-slip$ transitions and values of kinetic and static friction force increase, as to the frequency of
transitions, it decreases. At the pressure, corresponding to normal stress $n=n_4$, the $stick-slip$ mode is not realized. However,
the kinetic mode, corresponding to lubricant melting, does not set in and lubricant solidifies because of walls squeezing. As a result,
the lubricant can't melt, the friction force $F$ is high, it corresponds to solid-like lubricant and zero value of excess volume $f^2$,
as the squeezing of walls promotes the formation of long-range order of atoms in the lubricant. In the bottom figure panel, the
dependence is for the elevated temperature $T$. Here we have the kinetic mode of friction corresponding to a small value of the
friction force and to nonzero excess volume $f^2$. With pressure growth the $stick-slip$ mode is expected, and with a higher loading
the full solidification of the lubricant is to be observed the same as in the top panel of the figure for $n=n_4$. Thus, three modes
of friction have been revealed: 1) the kinetic mode in which the lubricant is always liquid-like; 2) the interrupted mode corresponding
to periodic melting/solidification; 3) the mode of dry friction characterized by larger value of the friction force and solid-like
lubricant structure. Similar modes are found also within the framework of stochastic model \cite{Filippov}.

\begin{figure}[!ht]
\hspace{-1cm} \centerline{\includegraphics[width=120mm]{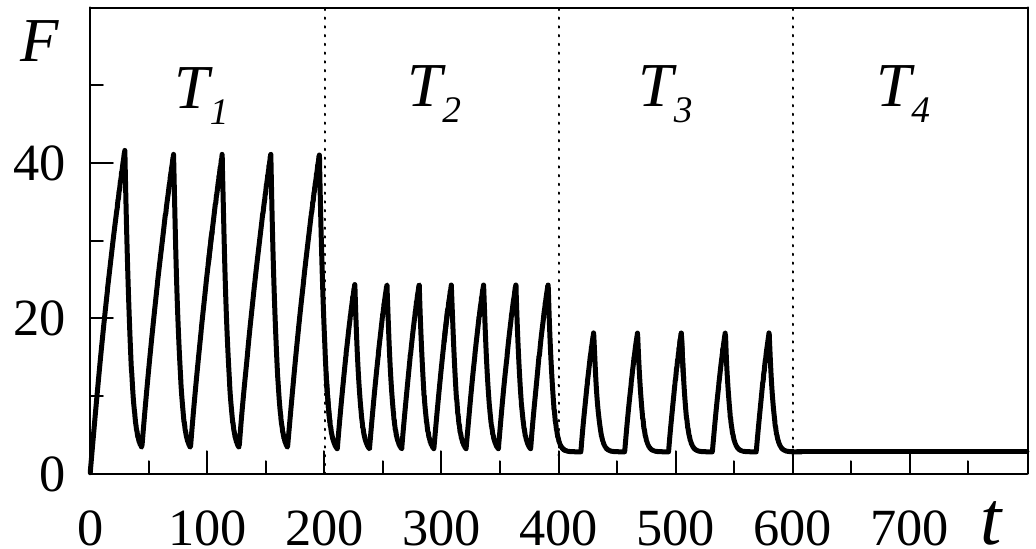}}
\caption{Dependence of the friction force $F$~(mN) on the time $t$~(s) for parameters of fig.~\ref{fig3}, velocity $V_0=3850$~nm, and lubricant temperatures $T_1=150$~K, $T_2=300$~K, $T_3=650$~K, $T_4=800$~K.} 
\label{fig6}
\end{figure}

In figure~\ref{fig6}, the dependence of friction force on lubricant temperature is shown.
Temperature rise leads to the decrease in the amplitude of friction force oscillations and the change of the frequency of phase
transitions. For $T=T_4$ the mode of sliding arises characterized by constant values of kinetic friction force and shear velocity of
the rubbing block. Thus, the temperature growth promotes lubricant melting. Experiments with a similar study of the temperature effect
are unknown to us and the given dependence is therefore predicting.

\section{Conclusion}\label{sec7}

The proposed theory allows to describe the effects observed at the melting of ultrathin lubricant film in a boundary friction mode.
The consistent consideration of thermodynamic and shear melting has been carried out. The dependences of friction force on shear
velocity and temperature, as well as on pressure upon surfaces are analyzed. At high lubricant temperature the shear melting is
realized at a smaller value of shear velocity (shear stresses), and with a still higher increase in temperature the lubricant melts
even at zero velocity of shear. The model takes the effect of temperature, shear melting, and pressure into consideration. These are
the major factors investigated experimentally.

Within the framework of the proposed theory a simple tribological system has been studied, and the time dependences of friction force
are obtained for the increasing shear velocity, pressure, and temperature. It is shown that in the system, in a wide range of
parameters, the experimentally observable intermittent mode of friction is realized. The obtained results  qualitatively coincide with
the known experimental data. Since the model is quantitative, it may be modified to describe concrete experiments.

The work has been done under financial support of the Fundamental Researches State Fund of Ukraine (Grant $\Phi$28/443-2009).

\end{document}